\documentstyle[12pt,preprint,aps]{revtex}

\tightenlines

\author{L. O. Manuel, A. E. Trumper and H. A. Ceccatto\\
Instituto de F\'{\i}sica Rosario, Consejo Nacional de \\
Investigaciones Cient\'{\i}ficas y T\'ecnicas and Universidad \\
Nacional de Rosario, Bvd. 27 de Febrero 210 Bis, 2000 Rosario, \\
Rep\'ublica Argentina}
\title{ Rotational invariance and order-parameter stiffness in 
frustrated quantum spin systems}

\begin{document}

\maketitle
\begin{abstract}
We compute, within the Schwinger-boson scheme, the Gaussian-fluctuation
corrections to the order-parameter stiffness of two frustrated quantum 
spin
systems: the triangular-lattice Heisenberg antiferromagnet and the 
$J_1\negthinspace-\negthinspace J_2$ model on the square lattice. For 
the triangular-lattice Heisenberg antiferromagnet we found that the 
corrections
weaken the stiffness, but the ground state of the system remains 
ordered in
the classical 120$^{\circ }$ spiral pattern. In the case of the 
$J_1\negthinspace-\negthinspace J_2$ model, with increasing frustration 
the stiffness is reduced until it vanishes, leaving a small window $0.53 
\lesssim \eta \lesssim 0.64$ where the system has no long-range magnetic
order. In addition, we discuss several methodological questions related 
to the Schwinger-boson approach. In particular, we show that the 
consideration
of finite clusters which require twisted boundary conditions to fit the
infinite-lattice magnetic order avoids the use of {\it ad hoc} factors to
correct the Schwinger-boson predictions.
\end{abstract}

\section{Introduction}

Two-dimensional (2D) frustrated quantum spin systems are a fascinating
subject that has been studied for more than 20 years now. Since the 
seminal work by Fazekas and Anderson on the $S=\frac 12$
triangular-lattice
Heisenberg antiferromagnet (TLHA),\cite{FA} the existence of 
non-magnetic
ground states in these systems has been strongly debated in the 
literature.
In the last years, this problem was somehow linked to more general 
questions
concerning spin-liquid states and their possible connections to 
high-$T_c$
superconductivity. In this context, the square-lattice Heisenberg
antiferromagnet with positive first- and second-neighbor interactions, 
the so-called $J_1\negthinspace -\negthinspace J_2$ model, has been much 
studied.\cite{12} Unlike the TLHA which is frustrated because of the 
lattice topology, in the $J_1\negthinspace-\negthinspace J_2$ model
frustration is due to the competing couplings.

Regarding the TLHA, although in general there is a growing conviction 
that this model displays the characteristic 120$^{\circ }$ spiral order 
in its
ground state (perhaps with an important reduction of the classical 
$\frac 12$
value),\cite{orden,Bernu,Claire} using different methods some authors 
found a
situation very close to a critical one or no magnetic long-range order 
at all.\cite{desorden,Leung} We have previously performed a study of the
TLHA
---including first- and second-neighbor interactions--- using the 
rotational
invariant Schwinger-boson approach in a mean-field (saddle-point)
approximation.\cite{SP} In particular, for the ground-state energy this
method showed a remarkable agreement with exact numerical results on 
finite
lattices (at the time of this work only results for the 12-site cluster 
were available). Furthermore, the local magnetization in the thermodynamc
limit
was predicted to be slightly larger than the (harmonic) spin-wave result.
More recently, new finite-size results\cite{Bernu,Leung} and 
$1/S$-corrected
spin-wave calculations\cite{Chubukov,Leche} became available, which lead 
us to reconsider this model. In this work we present the one-loop
corrections
to the saddle-point results obtained in \cite{SP}, along the lines 
developed
in \cite{stiff,letter}. We have computed the ground-state energy and the
spin-stiffness tensor in order to assess the existence or not of magnetic
order, and we have compared these results with exact values on finite
lattices and spin-wave results.

As for the $J_1\negthinspace-\negthinspace J_2$ model, in previous
publications \cite{stiff,letter} we have considered its behavior on 
finite
clusters that do not frustrate the short-range N\'eel and collinear 
orders
present in this system. In particular, we computed the stiffness $\rho $ 
to determine the range of $\eta =J_2/J_1$ where the magnetic order is
destroyed by the combined action of quantum fluctuations and frustration.
However, it turned out to be that the value of $\rho $ obtained at
saddle-point order was nearly a factor 2 smaller than exact results on 
small
clusters. This was confirmed by considering the large-$S$ limit of the
Schwinger-boson predictions, which differed from the classical results by
exactly this factor. The same happened, now for all values of $S$, in the
2-site problem, which can be worked out exactly. After these checks, 
in \cite
{stiff,letter} we simply added an {\it ad hoc} factor 2 to correct the
saddle-point result for $\rho $. Although the need for this kind of 
factors
has been stressed since the first works on Schwinger bosons,\cite{AA} 
their use is highly unsatisfactory.

In the course of the present study on clusters of the triangular lattice,
again the large-$S$ Schwinger-boson prediction for $\rho $ did not 
reproduce
the classical result. The situation was here even worse than for the 
square
lattice, since now the rotational invariant Schwinger-boson approach 
could
not produce concrete predictions for the stiffness parallel and
perpendicular to the plane where the order parameter spirals. In an 
attempt
to resolve this last problem, we followed the idea in \cite{Bernu,Leche} 
and
considered clusters with appropiate twisted boundary conditions to fit 
the
120$^{\circ }$ spiral order in the $xy$-plane. In this case, since the
boundary conditions break the rotational invariance explicitly, one is 
able
to compute the parallel and perpendicular stiffness separately. Moreover, 
we found that now the large-$S$ predictions for these quantities have the
correct behavior and no {\it ad hoc} factors are required. This 
unexpected
result prompted us to reinvestigate the 
$J_1\negthinspace-\negthinspace J_2$ model on clusters which require the
use of antiperiodic boundary conditions
to avoid frustrating the N\'eel and collinear orders. Again in this case 
the results have the expected large-$S$ behavior, and there is no need 
for correction factors. These findings point to a complex interplay
between
rotational invariance and the unphysical enlargement of Fock space in the
Schwinger-boson approach. Fortunately, working on appropiate clusters the
results for both the TLHA and the $J_1\negthinspace-\negthinspace J_2$
model behave consistently, as we show below.

\section{Order-parameter stiffness and the Schwinger-boson approach}

Let's consider a general Heisenberg model, 
\begin{equation}
\label{H} H=\frac 12 \sum_{\bf r,r^{\prime}} J({\bf
r-r^{\prime}%
}) \ {\vec S}_{{\bf r}}\cdot {\vec S}_{{\bf r^{\prime}}} 
\end{equation}
where ${\bf r,r^{\prime}}$\ indicate sites on a two-dimensional lattice. 
On
finite clusters generated by the translation vectors ${\bf T}_\alpha$ 
($\alpha=1,2$) we impose arbitrary boundary conditions 
${\vec S}_{{\bf r+T}_\alpha}={\cal R}_{\widehat{n}}
(\Phi _\alpha){\vec S}_{{\bf r}}$, where 
${\cal R}_{\widehat{n}}(\Phi _\alpha)$ is the matrix that rotates an 
angle $\Phi _\alpha$ around some axis $\widehat{n}$ (notice that we are 
using
boldface (arrows) for vectors in real (spin) space). By performing local
rotations ${\vec S}_{{\bf r}}\rightarrow {\cal R}_{\widehat{n}} 
(\theta _{{\bf r}}){\vec S}_{{\bf r}}$ of angle $\theta _{{\bf r}}= 
{\bf Q\cdot r}$, the Hamiltonian (\ref{H}) becomes 
\begin{equation}
\label{HR} 
H=\frac 12 
\sum_{\bf r,r^{\prime}} J({\bf r-r^{\prime}}) \left\{ ({\vec S}%
_{{\bf r}}\cdot \widehat{n})({\vec S}_{{\bf r^{\prime}}}\cdot \widehat{n}%
)+\cos \theta _{{\bf rr^{\prime}}}\left[ {\vec S}_{{\bf r}} 
\cdot {\vec S}_{%
{\bf r^{\prime}}}-({\vec S}_{{\bf r}}\cdot \widehat{n}) 
({\vec S}_{{\bf %
r^{\prime}}}\cdot \widehat{n})\right] + \sin \theta _{{\bf %
rr^{\prime}}}({\vec S}_{{\bf r}}\times {\vec S}_{{\bf r^{\prime}}})
\cdot \widehat{n}\right\} , 
\end{equation}
where $\theta _{{\bf rr^{\prime}}}={\bf Q\cdot (r^{\prime}-r)}$. In this
way, with the choice ${\bf Q\cdot T}_\alpha=\Phi _\alpha$ the boundary
conditions become the standard periodic ones 
${\vec S}_{{\bf r+T}_{\alpha}}={\vec S}_{{\bf r}}$.

We define the ($T=0$) stiffness tensor $\rho _{\widehat{n}}$ by 
\begin{equation}
\label{RHO}\rho _{\widehat{n}}^{\alpha \beta }=\left. 
\frac{\partial ^2E_{%
{\rm GS}}({\bf Q)}}{\partial \theta _\alpha \partial \theta _\beta }
\right|_{{\bf Q=}0}, 
\end{equation}
where $E_{{\rm GS}}$ is the ground-state energy {\it per spin} and 
$\theta _\alpha
={\bf Q\cdot e}_\alpha $ ($\alpha $$=1,2$) are the twist angles along the
basis vectors ${\bf e}_\alpha $. Then, by using second-order perturbation
theory it is simple to prove that 
$\rho _{\widehat{n}}^{\alpha \beta }=
{\rm T}_{\widehat{n}}^{\alpha \beta }
+{\rm J}_{\widehat{n}}^{\alpha \beta },$ where
\begin{equation}
\label{T}{\rm T}_{\widehat{n}}^{\alpha \beta }=\left\langle -\frac 12%
\sum_{\bf r} J({\bf r)}r^\alpha r^\beta \left[ {{%
\vec S_0\cdot \vec S_{{\bf r^{}}}-}(\vec S_0\cdot \widehat{n})
(\vec S_{{\bf r%
}}\cdot \widehat{n})}\right] \right\rangle _{{\rm GS}}, 
\end{equation}
and 
\begin{equation}
\label{J}{\rm J}_{\widehat{n}}^{\alpha \beta }=2\left\langle {\rm j}_{ 
\widehat{n}}^\alpha \,P\left( \frac 1{H-E_{{\rm GS}}}\right) 
P\,{\rm j}_{\widehat{n}}^\beta \right\rangle _{{\rm GS}}. 
\end{equation}
Here \thinspace {\rm j}$_{\widehat{n}}^\alpha =\frac 12\sum_{{\bf r}}J(%
{\bf r)}r^\alpha ({\vec S_0}\times {\vec S_{{\bf r}}})\cdot 
{{\widehat{n}%
,\,\ }}P=1-\left| {\rm GS\left\rangle \negthinspace
\right\langle GS}\right| ,$ and ${\bf r=}\sum_\alpha r^\alpha 
{\bf e}_\alpha $. For further use, it is convenient to consider also 
the rotational 
average $\overline{\rho }_{xy}$ of the stiffness tensor for 
$\widehat{n}$ in 
the $xy$-plane. With $\widehat{n}=(\sin \theta \cos \phi ,\sin \theta 
\sin \phi,\cos \theta )$ and $\theta={\frac \pi 2}$, using 
$\int_0^{2\pi }\frac{d\phi }{2\pi }n^in^j=\frac{%
\delta ^{ij}}2$ ($i,j=1,2$) one obtains 
$\overline{\rho }_{xy}=\frac 12(\rho _{\widehat{x}}
+\rho _{\widehat{y}})$.
Analogously, using $\int \frac{d\theta d\phi }{4\pi }n^in^j=
\frac{\delta
^{ij}}3$ ($i,j=1,3$), for the corresponding average in all directions 
in spin space one obtains $\overline{\rho }=\frac 13
(\rho _{\widehat{x}}+\rho _{\widehat{y}}+\rho _{\widehat{z}})$.

In the case of classical spins, the only contribution to the stiffness 
comes from the {\rm T}-term (\ref{T}). For quantum spins, besides 
the fact that
the calculations are much more complicated, there is also a very 
important
conceptual difference with the classical case. This is more clearly
explained for the simple square-lattice Heisenberg model, where from the
Lieb-Mattis theorem we know that the exact ground state must be an 
isotropic
singlet $S=0$ (there is numerical evidence\cite{Bernu} that this 
theorem ---in principle valid for bipartite lattice--- might be
conveniently
extended to the triangular and other frustrated lattices). Then, if we
approach the infinite-lattice behavior using a technique which respects 
the
rotational invariance ---like, for instance, the Lanczos method on small
clusters---, we only have access to the stiffness average 
$\overline{\rho }.$
On the other hand, if we use approximate methods that assume an
order-parameter symmetry breaking ---like ordinary spin-wave theory, 
for instance---, then both the parallel $\rho _{\Vert }$ and perpendicular
$\rho_{\bot}$ stiffness tensors can be computed. There is yet a third
possibility, the use of the Schwinger-boson technique, which keeps the
rotational invariance on finite lattices and develops a spontaneous
symmetry-breaking (through a Bose condensation of the Schwinger bosons) 
in
the thermodynamic limit. We will consider this last method since, in
addition to providing us with a fairly reliable computational tool, there
are several interesting methodological questions which can be discussed.

First, we represent spin operators in terms of Schwinger bosons:\cite{AA} 
${\vec S_i}\negthinspace =\negthinspace {\frac 12}{\bf a}_i^{\dagger }
.{\bf \vec \sigma }.{\bf a}_i$, where ${\bf a_i^{\dagger }}\negthinspace 
=\negthinspace (a_{i\uparrow }^{\dagger },a_{i\downarrow }^{\dagger })$ 
is a
bosonic spinor, ${\bf \vec \sigma }$ is the vector of Pauli matrices, 
and
there is a boson-number restriction $\sum_\sigma a_{i\sigma }^{\dagger
}a_{i\sigma }\negthinspace =\negthinspace 2S$ on each site. Thus, we can
write the partition function for the Hamiltonian (\ref{HR}) as a 
functional
integral over boson coherent states, which allows a saddle-point 
expansion
to be performed. Following the calculations in \cite{letter}, which we 
do not reproduce here, we obtain the fluctuation-corrected ground-state
energy
as the sum of the saddle-point and one-loop contributions, $E_{{\rm GS}}
({\bf Q})=E_0({\bf Q})+E_1({\bf Q}).$ We evaluate $E_{{\rm GS}}({\bf Q})$ 
on
finite lattices in order to avoid the infrared divergencies associated 
to Bose condensation, which signals the appearance of magnetic 
long-range order
in the ground state. Numerical differentiation of 
$E_{{\rm GS}}({\bf Q})$ according to (\ref{RHO}) gives finally the spin
stiffness.

In what follows we will use this method to study the order-parameter
stiffness of both the TLHA and the $J_1\negthinspace-\negthinspace J_2$
model.

\section{Triangular-lattice Heisenberg antiferromagnet}

For the triangular lattice, finite clusters with the spatial symmetries 
of the infinite lattice correspond to translation vectors 
${\bf T}_1=(n+m){\bf e}_1+m{\bf e}_2$, ${\bf T}_2=n{\bf e}_1+(n+m)
{\bf e}_2$, where ${\bf e}_1=(a,0)$, ${\bf e}_2
=(-\frac 12a,\frac{\sqrt{3}}2a)$ ($a$ is the lattice
spacing).\cite{Bernu} The number of sites in these diamond-shaped 
clusters
is given by $N=n^2+m^2+nm$. When $2n+m$ and $n-m$ are multiples of $3$,
which corresponds to clusters with $N=3p$ ($p$ integer), periodic 
boundary
conditions do not frustrate the 120$^{\circ }$ spiral order. Otherwise, 
$N=3p+1$ and one has to use twisted boundary conditions with angles 
$\frac 23\pi (2n+m)$ and $\frac 23\pi (n-m)$ along the ${\bf T}_1$ and
${\bf T}_2$ directions, respectively.

Let's consider first the (by far) easiest case of classical spins. 
Choosing the proper boundary conditions as explained above, the classical
energy is
minimized by a configuration ${\vec S}_{{\bf r}}=S(\cos \theta _{{\bf r}}
,\sin \theta _{{\bf r}},0)$. Here $\theta _{{\bf r}}
= {\bf Q}_{{\rm spiral}}^{\triangle} \cdot {\bf r}$, with the magnetic 
wavevector
${\bf Q}_{{\rm spiral}}^{\triangle }=(\frac{4\pi}{3a},0)$. Since we took 
the spins to lie
on the $xy$-plane, it is necessary to consider both the parallel 
stiffness $\rho _{\Vert} \equiv \rho _{\widehat{z}}$ under a twist 
around $\widehat{z}$,
and the perpendicular stiffness $\rho _{\bot} \equiv 
\overline{\rho}_{xy}$
for twists around $\widehat{n}$ versors lying on this plane. As mentioned
before, for classical spins the only contributions to $\rho _{\Vert},\ 
\rho_{\bot}$ and $\overline{\rho}$ come from the ${\rm T}$ 
term (\ref{T}), and
one always obtains the tensorial structure $\rho^{\alpha \beta }
=\frac 12 (1+\delta ^{\alpha \beta })\rho$ with $\rho _{\Vert}=JS^2a^2$,
$\rho_{\bot}=\frac 12\rho _{\Vert}$ and $\overline{\rho}=\frac 23
\rho _{\Vert}$.
These last two results are consecuences of the identities 
$\overline{\rho}_{xy}=\frac 12(\rho _{\widehat{x}}+\rho _{\widehat{y})}$ 
and $\overline{\rho}=\frac 13(\rho _{\widehat{x}}+\rho _{\widehat{y}}+
\rho _{\widehat{z}})$,
plus the fact that $\rho _{\widehat{x}}=\frac 16\rho _{\Vert}$ and 
$\rho _{\widehat{y}}=\frac 56\rho _{\Vert }$ for the classical spiral 
arrangement we choosed.

In order to correct these results by considering the quantum nature of 
the
spins we have followed the procedure discussed in the previous section. 
On
the $N=3p$ clusters with periodic boundary conditions our approach is
rotationally invariant and we only have access to $\overline{\rho }$. 
Since
now $\rho _{\Vert }$ and $\rho _{\bot }$ are unrelated, the theory 
cannot
produce concrete predictions for them from $\overline{\rho }$. On the 
other
hand, we found that the large-$S$ Schwinger-boson prediction for this
last quantity is exactly 4/3 smaller than the corresponding classical 
result. 
As mentioned in the introduction, in an attempt to resolve the former 
problem
we followed the idea in \cite{Bernu,Leche}, and considered the $N=3p+1$
clusters with the appropiate twisted boundary conditions to fit the 
120$^{\circ }$ spiral order in the $xy$-plane. In this case, since the 
boundary
conditions break the rotational invariance explicitly, one is able to
compute the parallel and perpendicular stiffness separately. Furthermore, 
we found that now the large-$S$ predictions have the correct behavior 
and no {\it ad hoc} factors are required.

In Fig. 1  we present the results for the energy of the $N=3p$
clusters, together with results from exact diagonalization 
studies\cite{Claire,orden}.   
As can be seen, for $N=12,36$, after the inclusion of Gaussian 
fluctuations
our theory gives predictions very close to the exact ones. However, for 
$N=21,27$ the agreement is not so good. This can be explained by noticing
that for odd number of sites the true ground state has total spin $S=1/2$,
while the approximate Schwinger-boson wave function is always a singlet
(rotational symmetries are broken only in the thermodynamic limit by the
boson condensate). The difference between even and odd $N$ is also 
apparent
in the exact results, since these do not line up according to the 
expected
(spin-wave) scaling. However, following \cite{Claire}, after 
substracting
the top inertial effects in the exact results for the odd-$N$ clusters, 
our results and these corrected values are in very good agreement (see
Fig. 1). This problem becomes less important for larger
clusters. The inset in Fig. 1 shows the thermodynamic-limit
extrapolation on large lattices of both the $N=3p$ and $N=3p+1$ types
(notice that for the $N=3p$ clusters the linear trend has a slope 
rather
different than what one would obtain from the consideration of the 
smallest
clusters). As discussed above, the $N=3p+1$ clusters require twisted
boundary conditions to avoid frustrating the spiral order. The main
consecuence of this is the absence of rotational symmetry in the 
ground
state, since the boundary conditions choose a plane of magnetization 
(the $xy$-plane in this case). Consequently, the scaling of the energy 
to the
thermodynamc limit has a quite different slope than the corresponding 
to $N=3p$ clusters, although, of course, the final value for the 
ground-state
energy {\it per site }$E_{{\rm GS}}\simeq -0.5533$ is the same in both
cases. This happens, as expected, with the results at saddle-point
order and also after the inclusion of Gaussian fluctuations.

The finite-size scaling of the parallel stiffness is shown in Fig. 2. 
In this figure we plot both the saddle-point and 
fluctuation-corrected results, which run almost parallel to each other. 
The
first-order spin-wave result of \cite{Leche} is also given for 
comparison.
Notice that, in spite of the different slope, the extrapolated values 
do not
differ much. As a further comparison, in Fig. 3 we plot the
renormalization factor between quantum and classical results for the
parallel stiffness on small lattices, as given by different techniques. 
As
can be seen from this figure, the introduction of 
Gaussian-fluctuation
corrections to the saddle-point Schwinger-boson results gives values 
closer
to the exact ones than those of first-order spin-wave theory.

\section{$J_1\negthinspace-\negthinspace J_2$ model}

For the square lattice, finite clusters of $N=\sqrt{n^2+m^2}\times 
\sqrt{n^2+m^2}$ sites correspond to translation vectors ${\bf T}_1
=n{\bf e}_1+m{\bf e}_2$, ${\bf T}_2=-m{\bf e}_1+n{\bf e}_2$, where 
${\bf e}_1=(a,0)$, ${\bf e}_2=(0,a)$. If $n,m$ are even, the periodic 
boundary conditions do not
frustrate the short-range N\'eel (${\bf Q}_{{\rm N\acute eel}}^{\Box
}=(\frac \pi a,\frac \pi a)$) and collinear (${\bf Q}_{{\rm coll}}^{\Box
}=(\frac \pi a,0)$) orders present for $\eta $$=J_2/J_1\ll 1$ and $\eta
\simeq 1$ respectively. For $n$ and/or $m$ odd one has to impose
antiperiodic boundary conditions to fit both magnetic patterns to the
cluster shape.

Let's assume that the classical energy is minimized by a spin 
configuration ${\vec S}_{{\bf r}}=S(\cos \theta _{{\bf r}},\sin 
\theta _{{\bf r}},0)$. For
simplicity we consider first the square lattice with nearest-neighbor
interactions only, so that $\theta _{{\bf r}}
={\bf Q}_{{\rm N\acute eel}}^{\Box} \cdot {\bf r}$. Then, from 
(\ref{T}) one obtains the tensorial
structure $\rho^{\alpha \beta}=\rho \delta ^{\alpha \beta}$, where, 
like for
the triangular lattice, $\rho _{\Vert}=JS^2a^2$, $\rho _{\bot}
=\frac 12\rho_{\Vert}$ and $\overline{\rho}=\frac 23\rho _{\Vert}$. 
Notice however that
for collinear spin arrangements like the N\'eel order there is properly 
no perpendicular stiffness. In fact, the value of $\rho _{\bot}$ is
obtained
from $\overline{\rho}_{xy}=\frac 12(\rho _{\widehat{x}}
+\rho _{\widehat{y}})$
and $\rho _{\widehat{x}}=0$, $\rho _{\widehat{y}}
= \rho _{\Vert}$ for
vectors pointing in the $\pm$$\widehat{x}$ directions. In the same way, 
the value of $\overline{\rho}$ is a consequence of the identity
$\overline{\rho}=\frac 13(\rho _{\widehat{x}}+\rho _{\widehat{y}} 
+\rho _{\widehat{z}}).$

In previous publications\cite{stiff,letter} we have considered finite
clusters with periodic boundary conditions. Then, we determined 
$\overline{\rho }$ and, from this quantity, $\rho _{\Vert }
=\frac 32\overline{\rho }$.
However, it turned out to be that the results obtained at 
saddle-point order
for $\rho _{\Vert }$ were nearly a factor 2 smaller (or, like in the
triangular lattice, the results for $\overline{\rho }$ nearly 
4/3 smaller)
than exact results on small clusters. As pointed out in the 
introduction,
this was confirmed by considering the large-$S$ limit of the 
Schwinger-boson
predictions, which differed from the classical results by exactly 
these
factors. Moreover, the same happened for all values of $S$ in the 
2-site
problem, which can be worked out exactly. After these checks, 
in \cite
{stiff,letter} we simply added an {\it ad hoc} factor 2 to correct the
saddle-point result for $\rho $$_{\Vert }$. The need for this kind of
factors is related to a subtle interplay between rotational invariance 
and
the relaxation of the local boson-number restriction, a claim that can 
be substantiated by the following argument. At saddle-point order the
energy $E_{\rm GS}({\bf Q})=\frac 12\sum_{{\bf r}}J({\bf r})
[B_{{\bf r}}^2({\bf Q})
-A_{{\bf r}}^2({\bf Q})]$, where $B_{{\bf r}}({\bf Q})$ and $A_{{\bf r}}
({\bf Q})$ are the order parameters for the spin-spin interaction in the 
ferromagnetic and antiferromagnetic channels respectively. At the same 
order 
in the calculations, the exact (local) constraints would force the 
identity 
$B_{{\bf r}}^2({\bf Q})+A_{{\bf r}}^2({\bf Q})\equiv S^2$, which shows
that both order parameters should contribute in the same way to the
stiffness (\ref{RHO}). Moreover, the same identity indicates that 
$(\frac{\partial A_{{\bf r}}}{\partial {\bf Q}})^2+(\frac{\partial 
B_{{\bf r}}}{\partial {\bf Q}})^2\equiv -(A_{{\bf r}}{\frac{\partial 
^2A_{{\bf r}}}{\partial {\bf Q}\partial {\bf Q}}+}B_{{\bf r}}
{\frac{\partial ^2B_{{\bf r}}}{\partial {\bf Q}\partial {\bf Q}})}$. 
However, when the constraints are
imposed on average the order parameters behave independently, the 
identity
is violated, and this last relationship is not fulfilled (the l.h.s.
vanishes). On the other hand, on clusters with antiperiodic boundary
conditions, which break the rotational invariance, this relationship 
between
derivatives does come right and leads to the correct value for the
stiffness, as shown below.

In Fig. 4 we plot the spin stiffness of the 
$J_1\negthinspace-\negthinspace J_2$ model on large clusters with 
antiperiodic boundary
conditions (we considered $\eta =0$ and 0.5 as examples). As expected, 
the
points line up according to the scaling behavior $\rho _{\Vert }
\sim N^{-1/2}$. This extrapolation to the thermodynamic limit for several 
values of $\eta 
$ produces the result shown in Fig. 5. There we see a small window
($\sim [0.53,0.60]$) where the stiffness vanishes and there is no 
magnetic
order (compare with the bottom panel in Fig. 3 of \cite{letter}). The 
inset
shows the scaling behavior of the point $\eta _{{\rm N\acute eel}}^c$ 
where
the N\'eel-order stiffness vanishes. Actually, the region without 
long-range
order extends at least up to $\eta \simeq 0.64$, since, as
shown in \cite{letter}, for $0.60\lesssim \eta \lesssim 0.64$ the
short-range antiferromagnetic order has lower energy than the 
long-range collinear one.

Let's consider also the scaling of the ground-state energy with $N$, 
$E_{{\rm GS}}^N({\bf Q}_{{\rm N\acute eel}}^{\Box })\simeq E_{{\rm GS}}
({\bf Q}_{{\rm N\acute eel}}^{\Box})-a/N^{3/2}$, along the series
of clusters with periodic and antiperiodic boundary conditions. The 
inset in Fig. 6 shows that both series extrapolate to
the same value ($E_{{\rm GS}}({\bf Q}_{{\rm N\acute eel}}^{\Box })
\simeq
-0.5015$ for $\eta=0.5$, within numerical errors), although with quite 
different slopes $a$.
It is of some interest to discuss the behavior of $a$, which, for 
periodic
boundary conditions, is proportional to the spin-wave velocity (or 
its
anisotropy average)\cite{Schulz}. Working on clusters with periodic 
boundary 
conditions, in our previous publication\cite{letter} we found
an anomalous behavior of $a$ near the point where the collinear order 
is melted (see Fig. 3 in \cite{letter}). In the present work, for
clusters with
antiperiodic boundary conditions, $a$ has a behavior more in line with 
{\it a priori} expectations (see Fig. 6).

\section{Conclusions}

We have considered the Gaussian-fluctuation corrections to the spin
stiffness of two frustrated quantum spin systems of interest: the 
TLHA and
the $J_1\negthinspace-\negthinspace J_2$ model. For the TLHA we found 
that
the fluctuation corrections weaken the stiffness, but the ground state 
of the system remains ordered in the classical 120$^{\circ }$ spiral 
state. In
the case of the $J_1\negthinspace-\negthinspace J_2$ model, with 
increasing
frustration the stiffness is reduced until it vanishes, leaving a 
small
window $0.53\lesssim \eta \lesssim 0.64$ where the system has no 
long-range magnetic order.

In the course of this investigation we discussed several methodological
questions related to the Schwinger-boson approach we used. In particular, 
we showed that the consideration of clusters which require twisted
boundary
conditions to fit the magnetic orders avoids the use of {\it ad hoc} 
factors
to correct the Schwinger-boson predictions. This fact points to a 
subtle
interplay between rotational invariance and the relaxation of local
constraints in this approach.

Finally, it is interesting to notice that for both the square and 
triangular
lattices with only nearest-neighbor interactions the 
fluctuation-corrected
spin stiffness scale almost parallel to the saddle-point results 
(see Figs. 2 and 4), although the corrections have different sign.
On the contrary, for large frustration in the square lattice the 
one-loop
result have a very different slope than the zero-order one, which 
ultimately
leads to disordering the system. This difference reflects the 
distinct nature of the order-parameter fluctuations in this highly
frustrated system.

\figure{FIG. 1: Ground-state energy {\it per site} $E_{\rm GS}^N$ for 
the TLHA as a function of the number of sites $N$ in the cluster. Open 
and full circles give the 
saddle-point and fluctuation-corrected results, respectively; pluses 
are exact numerical results from \cite{Leung} and crosses are the 
rotational corrected values from \cite{Claire}.
The line indicates the extrapolation to the thermodynamic 
limit. Inset: Extrapolation to the infinite
lattice along the series of clusters with periodic (full circles) and 
twisted (full squares) boundary conditions.}

\figure{FIG. 2: Parallel stiffness $\rho_{\Vert}$ for the TLHA as 
a function of
the number of sites $N$ in the cluster. Open and full circles give the 
saddle-point and fluctuation-corrected results, respectively. The full 
lines indicate the 
thermodynamic-limit extrapolations and the dashed line is the 
first-order 
spin-wave result of \cite{Leche}.}

\figure{FIG. 3: The renormalizaton factor 
$Z_{\Vert}= \rho_{\Vert}/\rho_{\Vert}^{\rm Cl}$ for the TLHA on 
small clusters. Open and full circles correspond to saddle-point (SP) 
and
fluctuation-corrected (FL) results, respectively. The exact values (EX)  
and first-order spin-wave (SW) results, indicated by crosses and open
squares respectively, are taken from \cite{Leche}.}

\figure{FIG. 4: Spin stiffness $\rho_{\Vert}$ for the 
$J_1\negthinspace-\negthinspace J_2$ model as a
function of the number of sites $N$ in the cluster. Top panel: $\eta=0$; 
bottom panel: $\eta=0.5$. Open and full circles give the saddle-point and 
fluctuation-corrected results, respectively. The lines indicate the 
thermodynamic-limit extrapolation.}

\figure{FIG. 5: Spin stiffness $\rho_{\Vert}$ for the 
$J_1\negthinspace-\negthinspace J_2$ model 
extrapolated to the thermodynamic limit. Dashed and full lines 
correspond
to saddle-point and fluctuation-corrected results, respectively. 
Inset: Extrapolation of the critical value $\eta_{\rm c}
=J_{2 {\rm c}}/J_1$
where the N\'eel-order stiffness vanishes.}

\figure{FIG. 6: Slope $a$ of the scaling $E_{\rm GS}^N$ vs. 
$N^{-3/2}$ on
clusters with antiperiodic boundary conditions. The inset shows the 
scaling of the ground-state energy $E_{\rm GS}^N$ on clusters with 
periodic 
(full circles) and antiperiodic (full squares) boundary conditions for
$\eta=0.5$.}

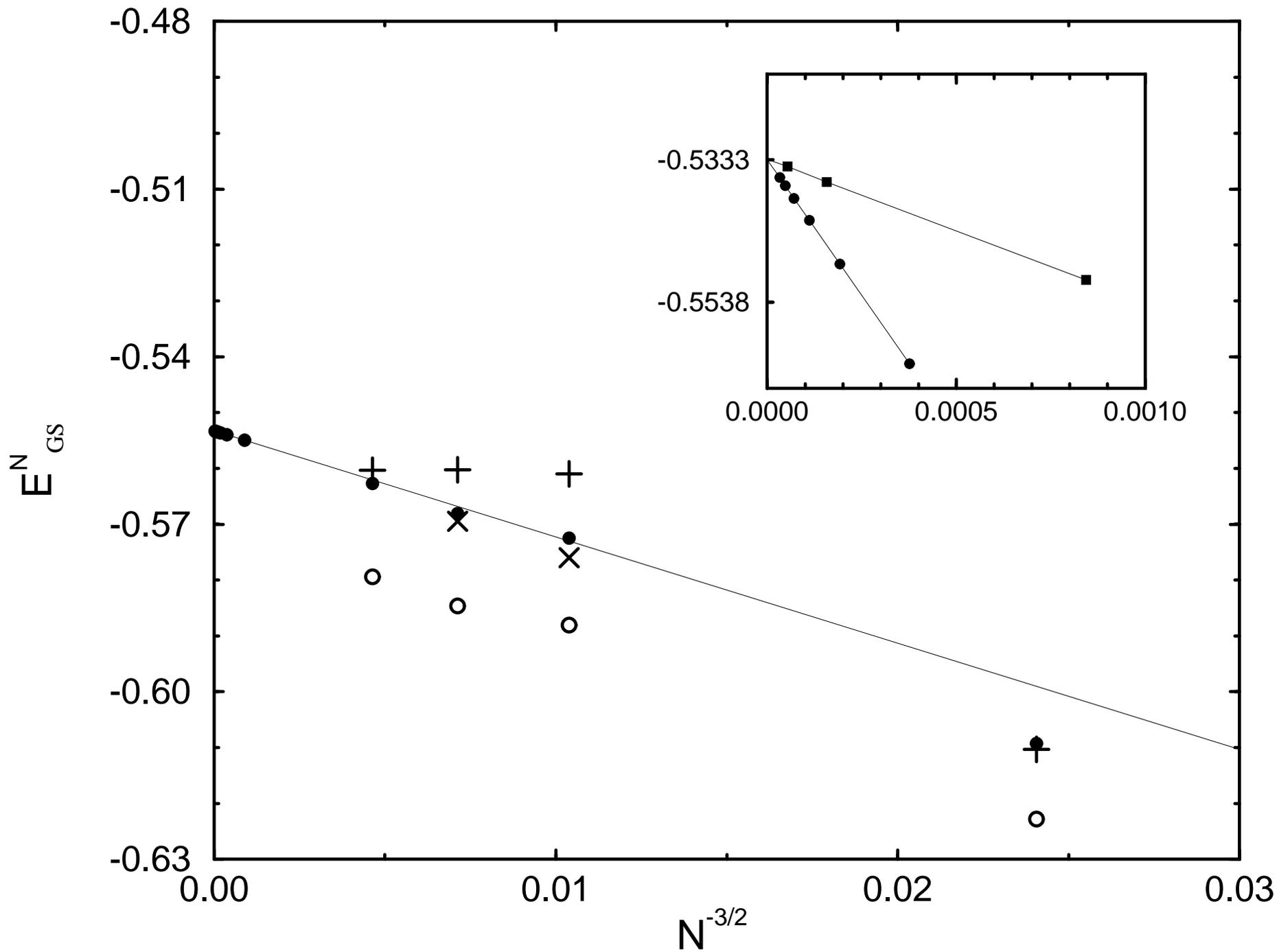
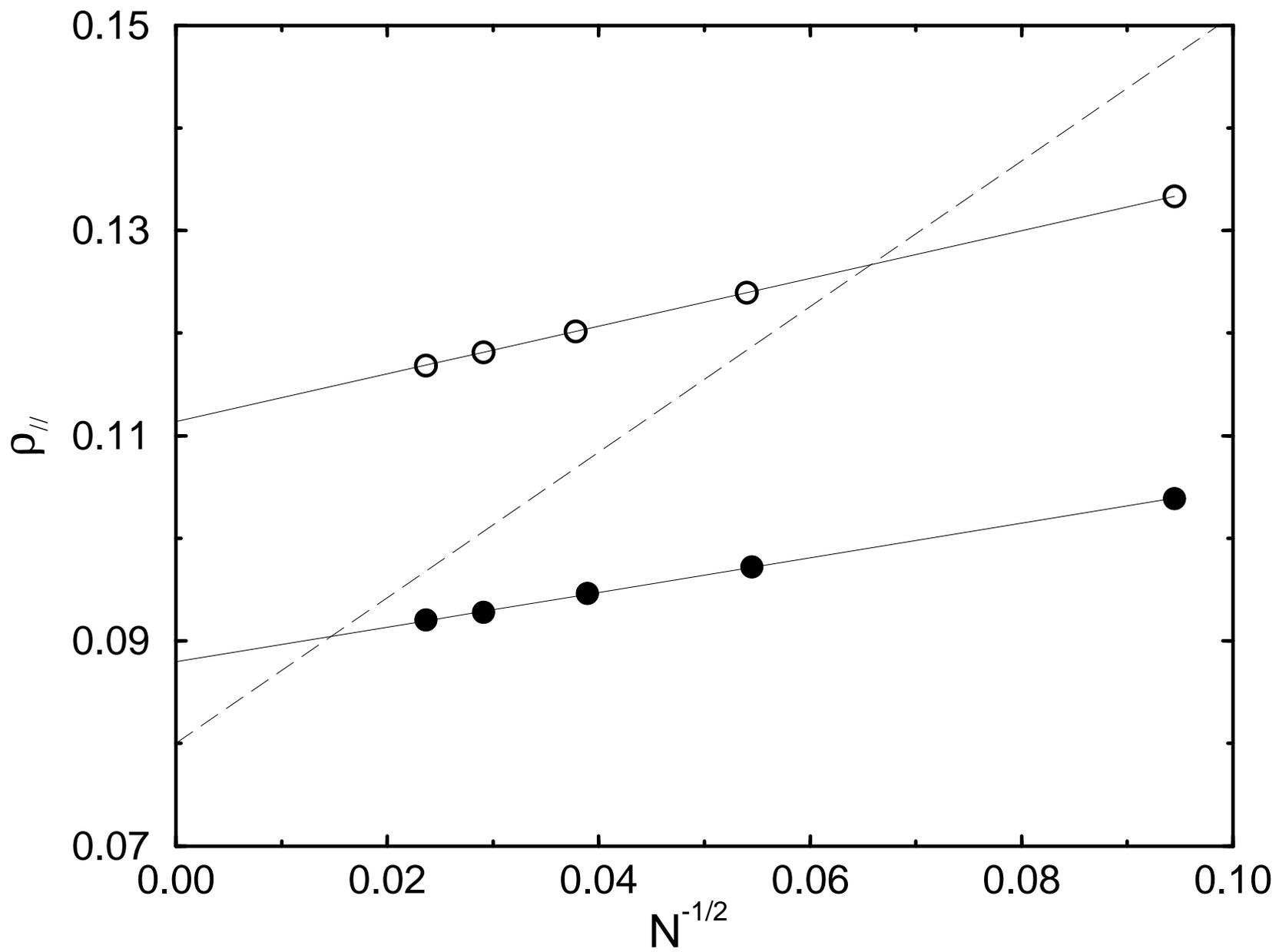
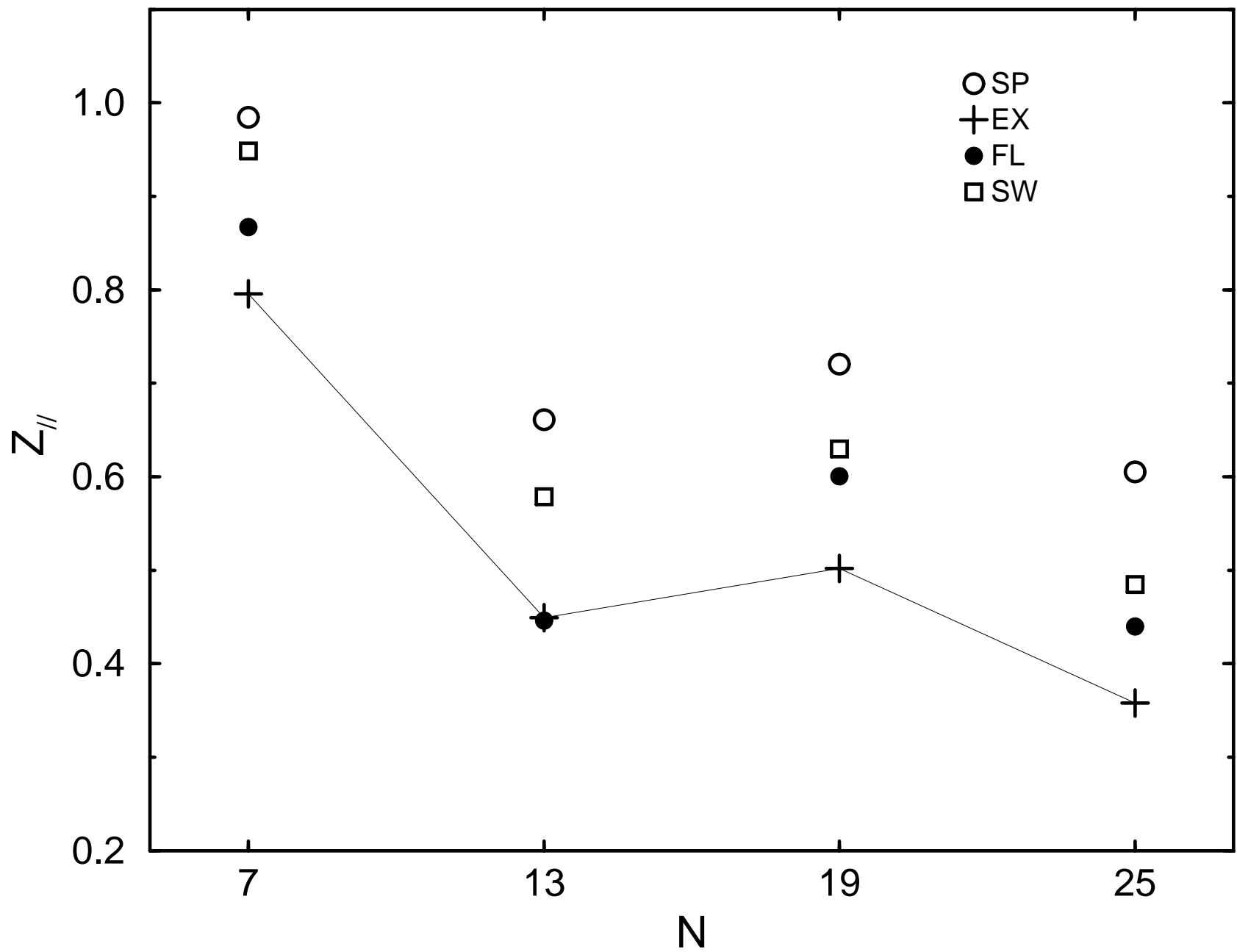
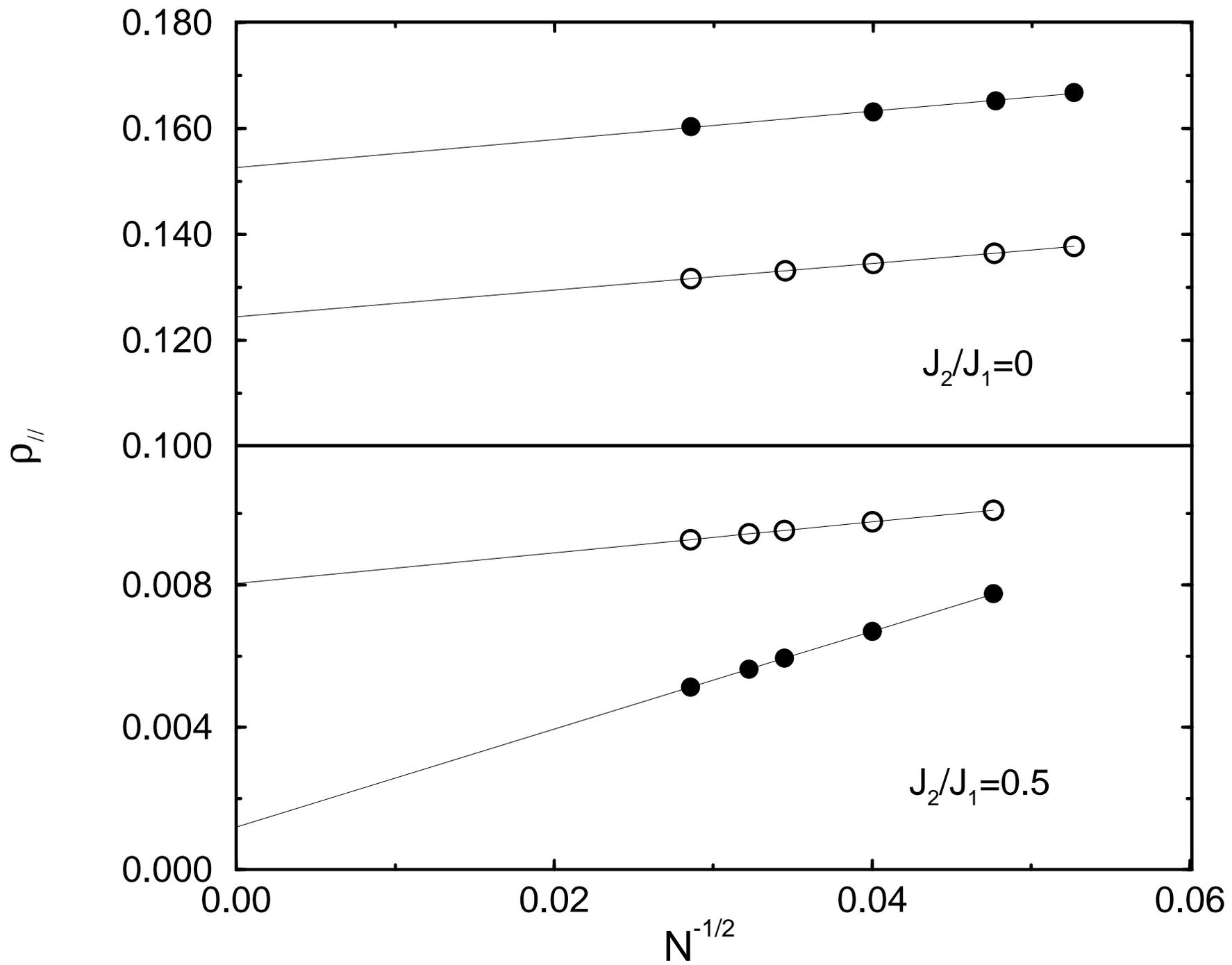
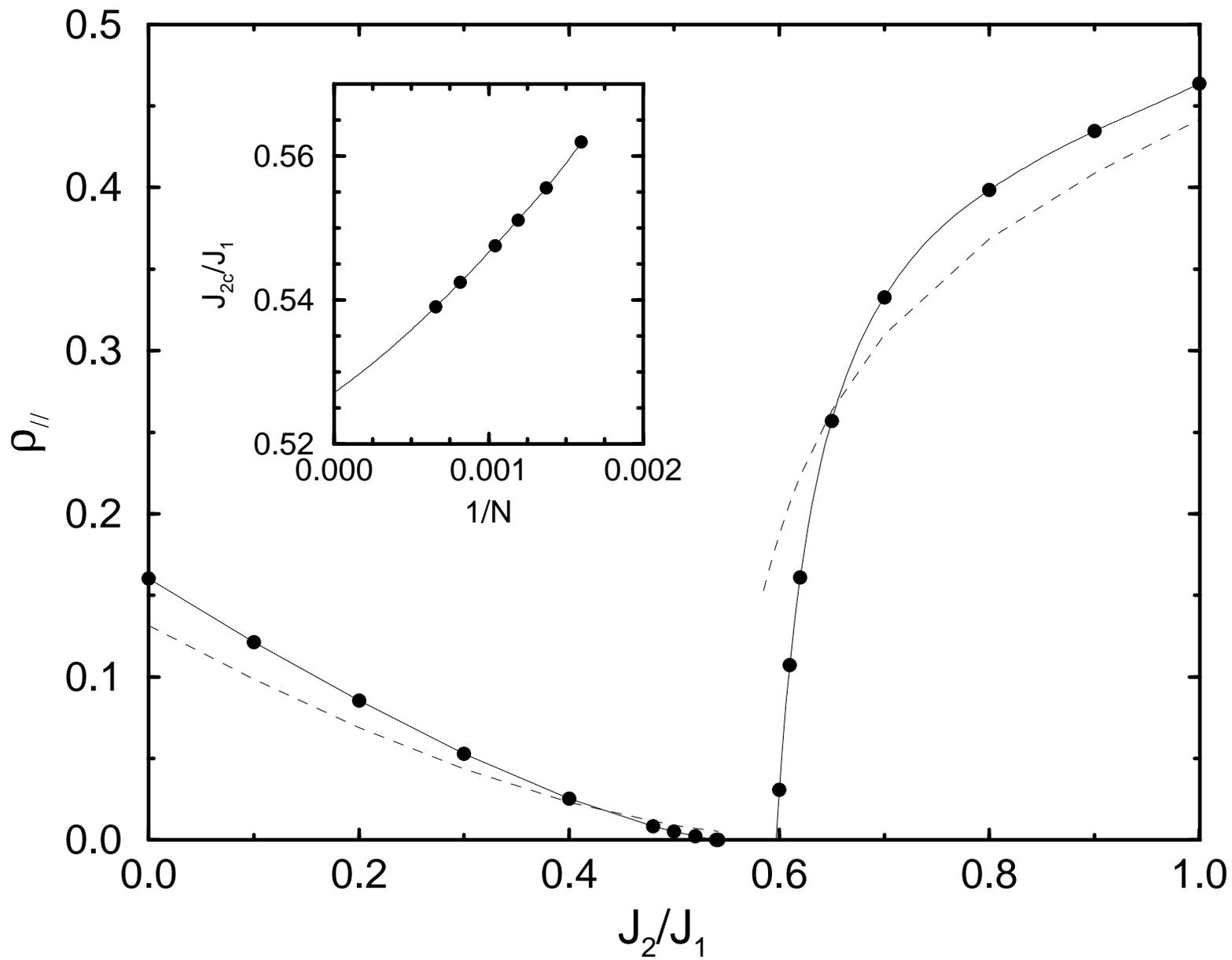
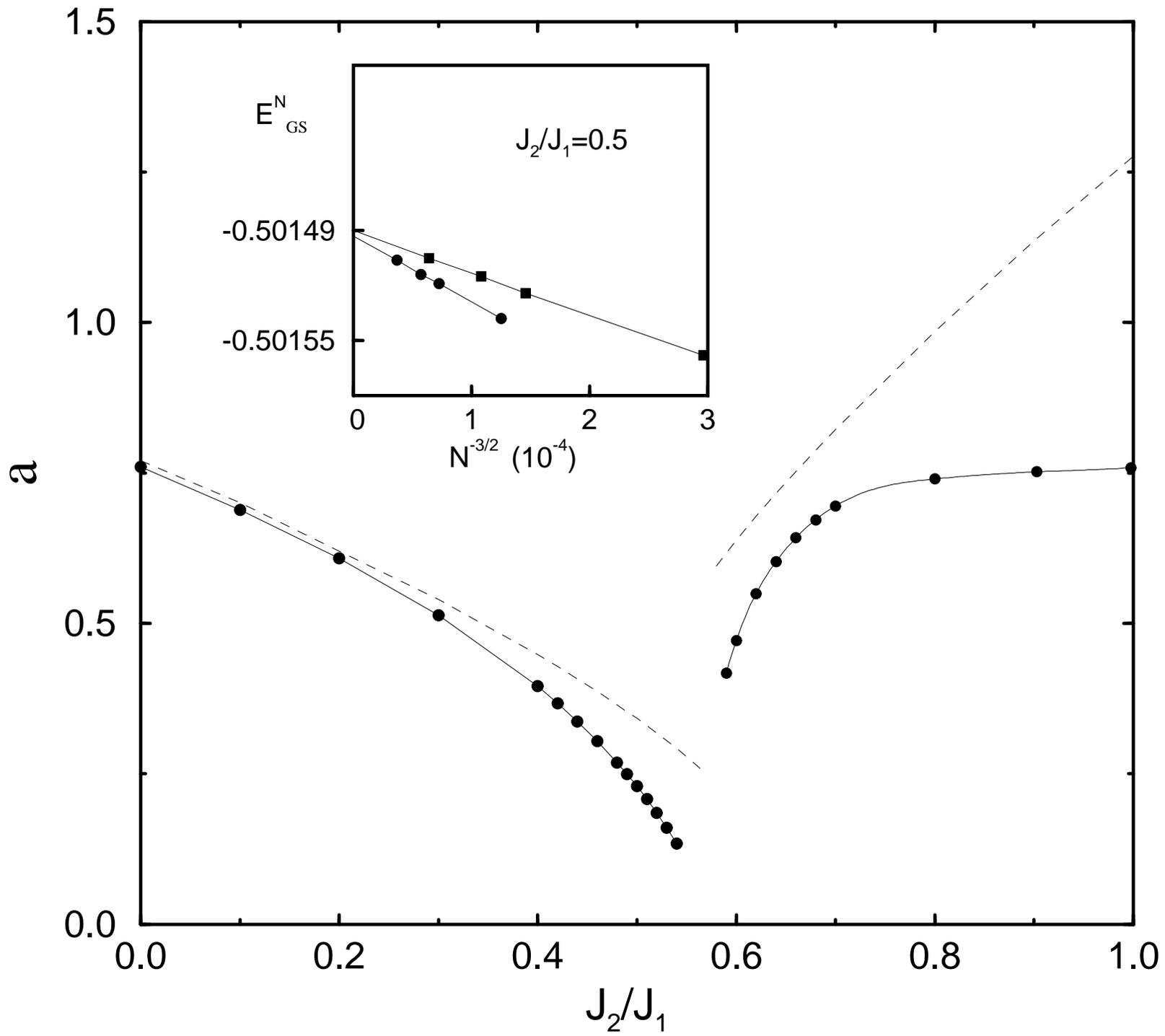
\end{document}